\documentclass{elsart5p}

\usepackage{epsfig}
\usepackage{amssymb}
\usepackage{graphicx}
\usepackage{subfigure}
\usepackage{here}

\def\figurename{\bf{Fig.\space}}

\begin{document}

\begin{frontmatter}

\title{Efficient ion blocking in gaseous detectors and its application to gas-avalanche photomultipliers sensitive in the visible-light range}

\author[w]{A. Lyashenko\corauthref{cor}},
\corauth[cor]{Corresponding author: tel. +972-8-934-2064, fax
+972-8-934-2611, E-mail: alexey.lyashenko@weizmann.ac.il}
\author[w]{A. Breskin},
\author[w]{R. Chechik}
\author[p1]{J.M.F. dos Santos},
\author[p1]{F.D.  Amaro} and
\author[p1,p2]{J.F.C.A. Veloso}
\address[w]{Department of Particle Physics, The Weizmann Institute of Science, 76100 Rehovot, Israel}
\address[p1]{Physics Department, University of Coimbra, 3004-516 Coimbra,  Portugal}
\address[p2]{Physics Department, University of Aveiro, Campus Universit\'ario de Santiago, 3810-193 Aveiro, Portugal}

\begin{abstract}

A novel concept for ion blocking in gas-avalanche detectors was
developed, comprising cascaded micro-hole electron multipliers with
patterned electrodes for ion defocusing. This leads to ion blocking
at the $10^{-4}$ level, in DC mode, in operation conditions adequate
for TPCs and for gaseous photomultipliers. The concept was validated
in a cascaded visible-sensitive gas avalanche photomultiplier
operating at atmospheric pressure of Ar/CH$_{4}$ (95/5) with a
bi-alkali photocathode. While in previous works high gain, in excess
of $10^{5}$, was reached only in a pulse-gated cascaded-GEM gaseous
photomultiplier, the present device yielded, for the first time,
similar gain in DC mode. We describe shortly the physical processes
involved in the charge transport within gaseous photomultipliers and
the ion blocking method. We present results of ion backflow fraction
and of electron multiplication in cascaded patterned-electrode
gaseous photomultiplier with K-Cs-Sb, Na-K-Sb and Cs-Sb
visible-sensitive photocathodes, operated in DC mode.

\end{abstract}

\end{frontmatter}

\section{Introduction}

Controlling the back-flow of ions generated in gas avalanches has
important consequences on the operation and properties of gaseous
detectors. Avalanche ions induce space charge effects that limit the
gain, counting-rate capability and localization properties of
tracking detectors. The ion impact on photocathodes (PC) of gaseous
photomultipliers (GPM) causes their permanent damage
\cite{breskin:05,singh:00}; more seriously, it induces the emission
of secondary electrons which, in turn, cause avalanche divergence,
deterioration of timing and localization information and, most
importantly, severe gain limits. All these consequences are included
in the term Ion Feedback effects. GPMs sensitive in the UV range,
with CsI PCs, do suffer some ion induced PC damage \cite{singh:00},
but the ion-induced secondary electron emission probability is very
low and does not limit their operation, even at high gain. There are
numerous large-area CsI-GPMs, presently operating or under
construction, in many particle-physics experiments, e.g. COMPASS
\cite{ketzer:02} and ALICE \cite{piuz:99} at CERN and PHENIX
\cite{fraenkel:05} at BNL.

The ion feedback effects are particularly problematic in
visible-sensitive GPMs, due to the high electron emission
probability of bi-alkali and other PCs sensitive in the visible
spectral range. The reader is referred to \cite{lyashenko:07} for an
extended discussion on this subject and references to recent works
dealing with methods of ion back-flow reduction.

The only method for significantly reducing the ion back-flow
fraction (IBF) and reaching high-gain operation in visible-sensitive
GPMs has been, so far, their operation in a gated mode
\cite{breskin:05}. Our goal has been to find methods for efficiently
reducing the IBF to permit the operation of visible-sensitive GPMs
in DC mode with single-photon sensitivity. Obviously, large tracking
TPCs will also benefit, as low IBF values would permit their stable
DC operation.

In the present article, we report on our recent results on blocking
of ion back-flow in cascaded micro-hole multipliers comprising GEMs
and other patterned electrodes. We discuss the ion-induced secondary
electron emission and provide solutions that permit, for the first
time, the operation of visible-sensitive GPMs in DC mode, with gain
of 10$^{5}$.

\section{Requirements for stable operation of visible-sensitive
GPMs}

\subsection{General consideration on Ion back-flow and Ion feedback effects.}

\begin{figure}[!h]
\begin{center}
\renewcommand{\thesubfigure}{\thefigure\alph{subfigure}}
\makeatletter
\renewcommand{\@thesubfigure}{\Large(\alph{subfigure})}
\renewcommand{\p@subfigure}{Fig.\space}
\renewcommand{\p@figure}{Fig.\space}
\makeatother %
\subfigure [] [] 
{
    \label{fig:1:a}
    \includegraphics[width=7cm]{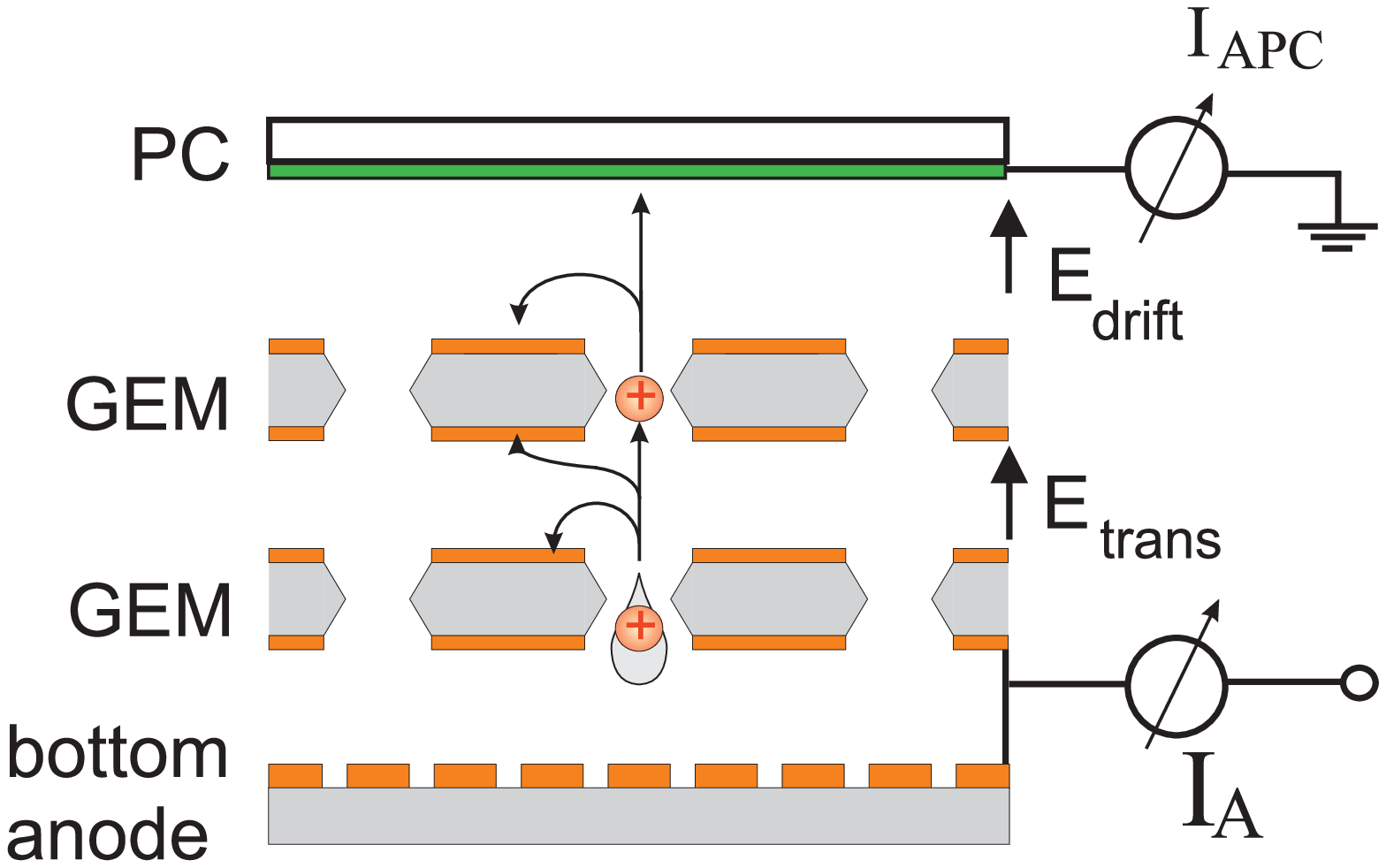}
}\hspace{0cm} %
\subfigure [] [] 
{
    \label{fig:1:b}
    \includegraphics[width=7cm]{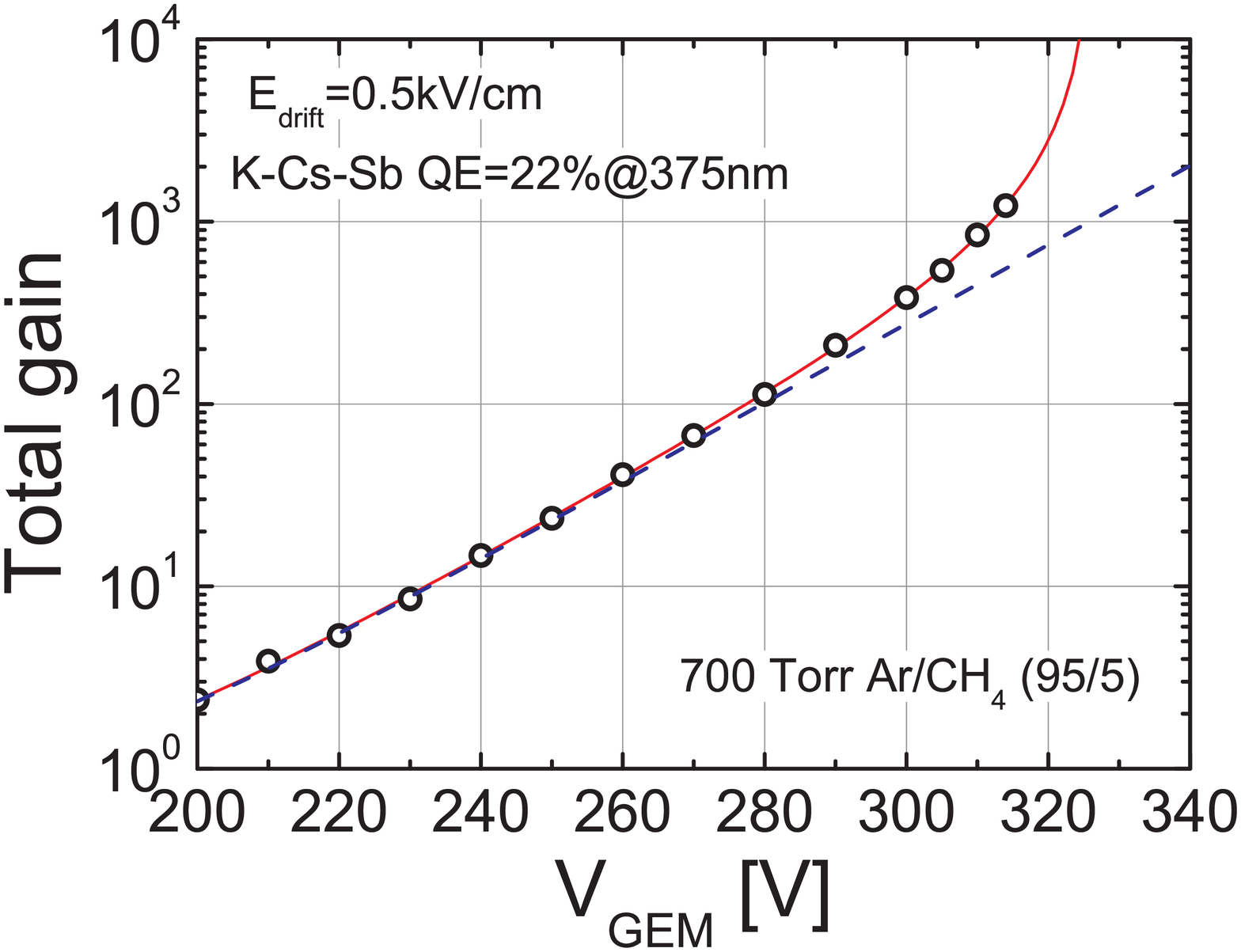}
}%
\caption {(a) A double-GEM GPM coupled to a semitransparent
photocathode; (b) gain-voltage characteristics measured in this GPM
(see conditions in the figure, QE refers to vacuum) with CsI
(dashed) and K-Cs-Sb (open circles) photocathodes. The divergence
from exponential with K-Cs-Sb is due to ion feedback.}
\label{fig:1}
\end{center}
\end{figure}

While cascaded micro-hole multipliers, with their significant
optical "opacity", efficiently block avalanche-photon feedback
\cite{moermann:04}, they are less efficient in blocking the
back-flow of avalanche ions. The latter, originating from each
avalanche stage in the cascaded multiplier, drift back to the GPMs'
PC following the device field lines, and a major fraction of them
follow the same paths (in opposite direction) of the initial
photoelectrons and their successive avalanche electrons
\cite{lyashenko:07}. When impinging on the PC's surface they release
secondary electrons; the latter initiate secondary avalanches, known
as ion-feedback, which, by positive feedback mechanism diverge the
proportional avalanche multiplication into discharge
\cite{moermann:thes}. An example of ion-feedback effect measured in
a double-GEM multiplier with K-Cs-Sb visible-sensitive PC operating
in Ar/CH$_{4}$ (95/5) at 700 Torr (\ref{fig:1:a}), is the deviation
of the gain-voltage curve from exponential (\ref{fig:1:b}). The
measured gain, G$_{meas}$, contains contributions from ion feedback
and is described by:

\begin{equation}
\label{eq:Gmeas}
\textrm{G}_{meas}=\frac{\textrm{G}}{1-\gamma_{+}\cdot\textrm{IBF}\cdot\varepsilon_{extr}\cdot\textrm{G}}
\end{equation}

where $\gamma_{+}$ is the \textbf{ion induced secondary emission
probability} or the ion feedback probability, IBF is the ion
back-flow fraction namely the fraction of ions, from all avalanche
stages of the multiplier, flowing back to the PC (or to the drift
region of a tracking detector), $\varepsilon_{extr}$ is the
efficiency of extracting secondary electrons from the PC into the
gas and G is the multiplier's gain without ion feedback. To avoid
avalanche divergence into a spark, the above formula should fulfill
$\gamma_{+}\cdot\textrm{IBF}\cdot\varepsilon_{extr}\cdot\textrm{G}<1$.
Therefore, a gain of 10$^5$, required for good single-photon
sensitivity in GPMs, implies
$\gamma_{+}\cdot\textrm{IBF}\cdot\varepsilon_{extr}<10^{-5}$.

\subsection{Measurement of $\gamma_{+}$ and $\varepsilon_{extr}$}

The extraction efficiency $\varepsilon_{extr}$ of secondary
electrons is the fraction of electrons emitted from the PC and not
scattered back (by collisions with gas molecules) into it
\cite{buzulutskov:00,escada:07}. $\varepsilon_{extr}$ depends on the
kinetic energy distribution of the electrons leaving the PC, which
has not yet been measured. The theoretical calculations of energy
distribution and extraction efficiency $\varepsilon_{extr}$ of
ion-induced secondary electrons from PCs are rather complex and
differ from those of photon-induced ones \cite{hagstrum:61}. Such
calculations are presently under way, in cooperation with T. Dias of
Coimbra University, and will be the subject of a future publication.

In the absence of any knowledge of $\varepsilon_{extr}$, we have
chosen to use $\gamma_{+}^{eff}=\gamma_{+}\cdot\varepsilon_{extr}$;
the latter can be extracted from the experimental GPM's gain curve
and its deviation from exponential line. $\gamma_{+}^{eff}$ is
defined as the effective ion induced secondary emission probability
or the effective ion feedback probability. It was extracted from
fitting the experimental gain curve G$_{meas}$ by equation
(\ref{eq:Gmeas}) (solid line in \ref{fig:1:b}). A significant
deviation from the exponential gain-voltage characteristic (dashed
line in \ref{fig:1:b}) is observed with the bi-alkali PC already at
low gain. The IBF and G as a function of GEM voltage were measured
in the same detector (geometry, gas and voltages), with a CsI PC,
ensuring no ion feedback. The drift field between the PC and the top
face of the first GEM was kept constant at 0.5kV/cm throughout the
entire measurements. In GPMs it provides about 60\% extraction
efficiency of photoelectrons in this gas \cite{moermann:thes}. The
gain-voltage characteristics, like the one shown in \ref{fig:1:b},
were measured for K-Cs-Sb, Na-K-Sb and Cs-Sb PCs; they yielded
$\gamma_{+}^{eff}$ values of $\sim$3$\cdot$10$^{-2}$ for all these
photocathodes. This study and the results will be discussed in more
detail elsewhere.

\subsection{Requirements for IBF}

Establishing the effective ion feedback value $\gamma_{+}^{eff}$
with the visible sensitive PCs under investigation, we can set the
limits on the IBF value needed for stable DC operation of
visible-sensitive GPMs at a gain of $10^{5}$.  Requiring
\\$\gamma_{+}^{eff}\cdot\textrm{IBF}<10^{-5}$, and using the
estimated value \\$\gamma_{+}^{eff}=\sim3\cdot10^{-2}$, the IBF
value should be $<3.3\cdot10^{-4}$.

\section{IBF reduction in cascaded micro-hole multiplier structures}

A straight forward way to reduce the IBF is by lowering the drift
field, since IBF decreases linearly with the drift field
\cite{bondar:03}. However, in GPMs the drift field could not be too
low because it controls, the photoelectron extraction into the gas,
; drift field values of the order of 0.5kV/cm \cite{moermann:thes}
were generally applied in our GPMs filled with Ar/CH$_{4}$ (95/5).

\begin{figure}[!h]
\begin{center}
\renewcommand{\thesubfigure}{\thefigure\alph{subfigure}}
\makeatletter
\renewcommand{\@thesubfigure}{\Large(\alph{subfigure})}
\renewcommand{\p@subfigure}{Fig.\space}
\renewcommand{\p@figure}{Fig.\space}
\makeatother %
\subfigure [] [] 
{
    \label{fig:2:a}
    \includegraphics[width=7cm]{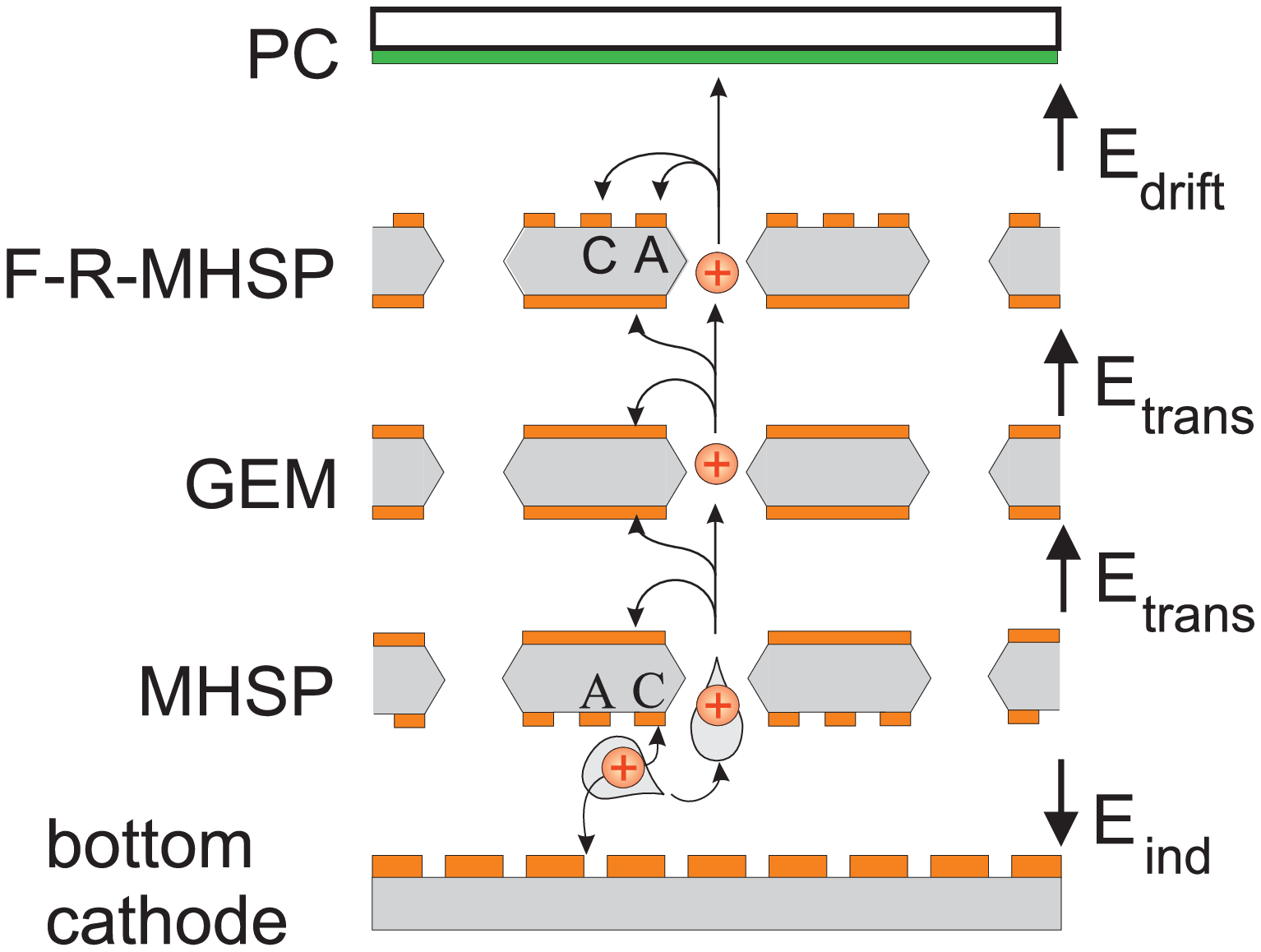}
}\hspace{0cm} %
\subfigure [] [] 
{
    \label{fig:2:b}
    \includegraphics[width=7cm]{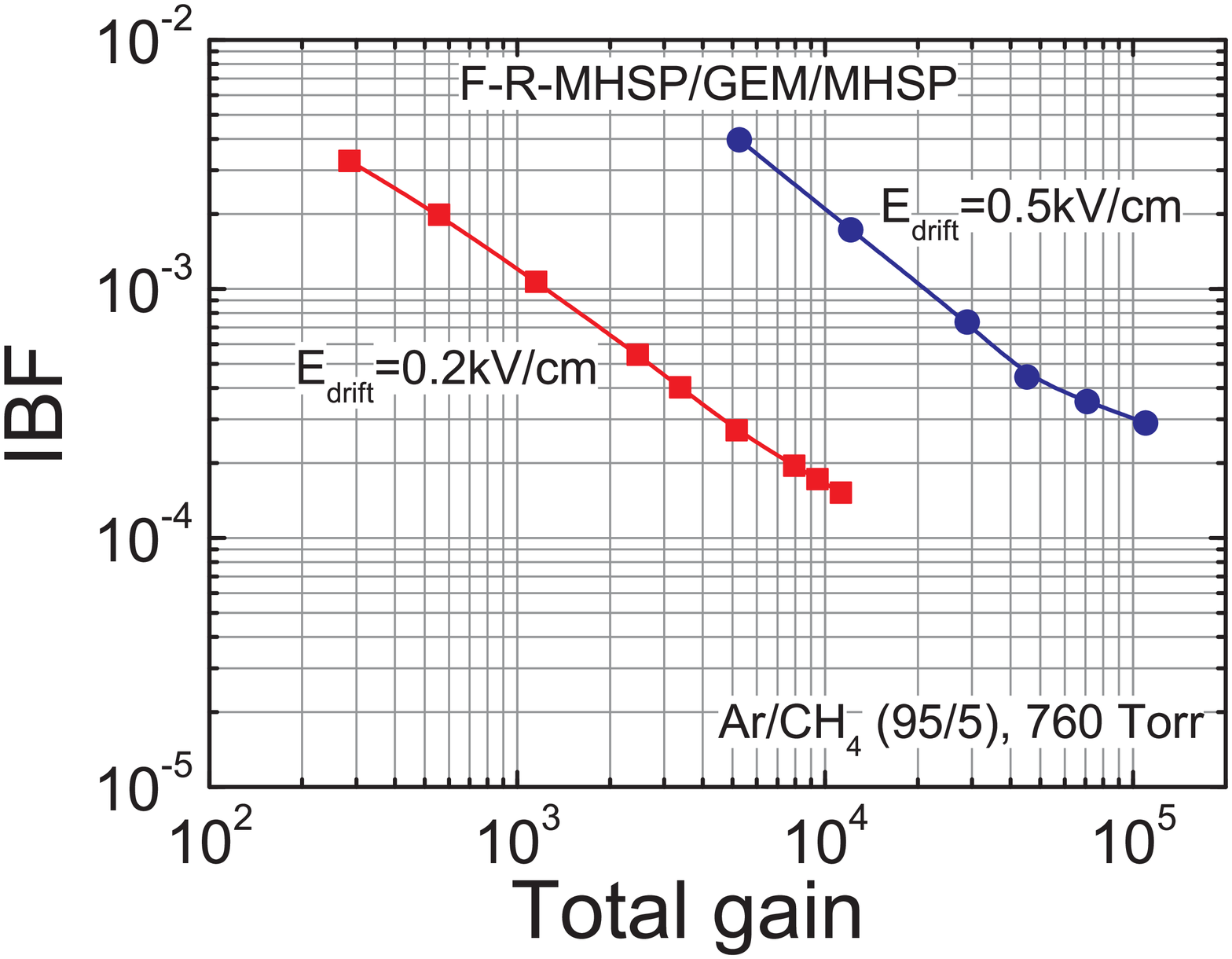}
}%
\caption {(a) Scheme of the cascaded F-R-MHSP/GEM/MHSP multiplier
coupled to a semi-transparent photocathode; possible avalanche-ion
paths are shown. (b) The IBF in correlation with the total gain of
this GPM plotted for drift fields of 0.2 kV/cm (TPC conditions,
squares) and 0.5 kV/cm (GPM conditions, circles).}
\label{fig:2}
\end{center}
\end{figure}

A more detailed report on a comprehensive IBF reduction, study,
carried out with a variety of cascaded micro-hole multipliers, can
be found elsewhere \cite{lyashenko:07}. All GPM detectors
investigated had active areas of 30x30$\textrm{mm}^2$; they were
irradiated over a surface of 15mm in diameter, at photon fluxes of
about $10^6$ $\frac{photons}{sec\cdot{cm^2}}$. The core outcome of
this study is that very low IBF values can be obtained with cascades
combining GEMs and other patterned electrodes derived from the
Micro-Hole \& Strip Plates (MHSP) \cite{veloso:00}. The latter
comprised GEM-like holes and additional patterned strips on one of
their faces, and they were operated in different modes regarding the
voltages and orientation schemes \cite{lyashenko:07}: MHSP,
reversed-MHSP (R-MHSP) and flipped-reversed-MHSP (F-R-MHSP). These
schemes aim at reducing the IBF by diverting the ions and trapping
them on the strips patterned on the surface. While the MHSP, placed
at the end of the cascade, can divert and trap only part of the ions
generated within its own avalanche stage, the other two types of
electrodes can divert ions created in successive multiplying
elements; therefore their incorporation in the cascade yielded
better results. In a cascade comprising F-R-MHSP followed by GEM and
MHSP (\ref{fig:2:a}), IBF values as low as $3\cdot10^{-4}$ were
recorded \cite{lyashenko:07}; this fulfills our requirement for
stable DC operation at a gain of 10$^{5}$ with visible-sensitive PCs
(see \ref{fig:2:b}). We varied the inter-strip voltage on the bottom
MHSP, to vary the total gain of the detector.

\begin{figure} [h]
  \begin{center}
  \makeatletter
    \renewcommand{\p@figure}{Fig.\space}
  \makeatother
    \epsfig{file=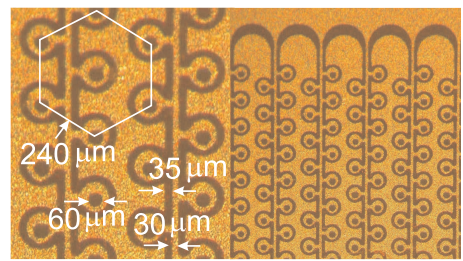, width=6cm}
    \caption{A microscope photographs of one face of a "Cobra" micro-hole electrode with dimensions given in the figure. The other face is identical to a GEM.}
    \label{fig:3}
  \end{center}
\end{figure}

\begin{figure}[!h]
\begin{center}
\renewcommand{\thesubfigure}{\thefigure\alph{subfigure}}
\makeatletter
\renewcommand{\@thesubfigure}{\Large(\alph{subfigure})}
\renewcommand{\p@subfigure}{Fig.\space}
\renewcommand{\p@figure}{Fig.\space}
\makeatother %
\subfigure [] [] 
{
    \label{fig:4:a}
    \includegraphics[width=7cm]{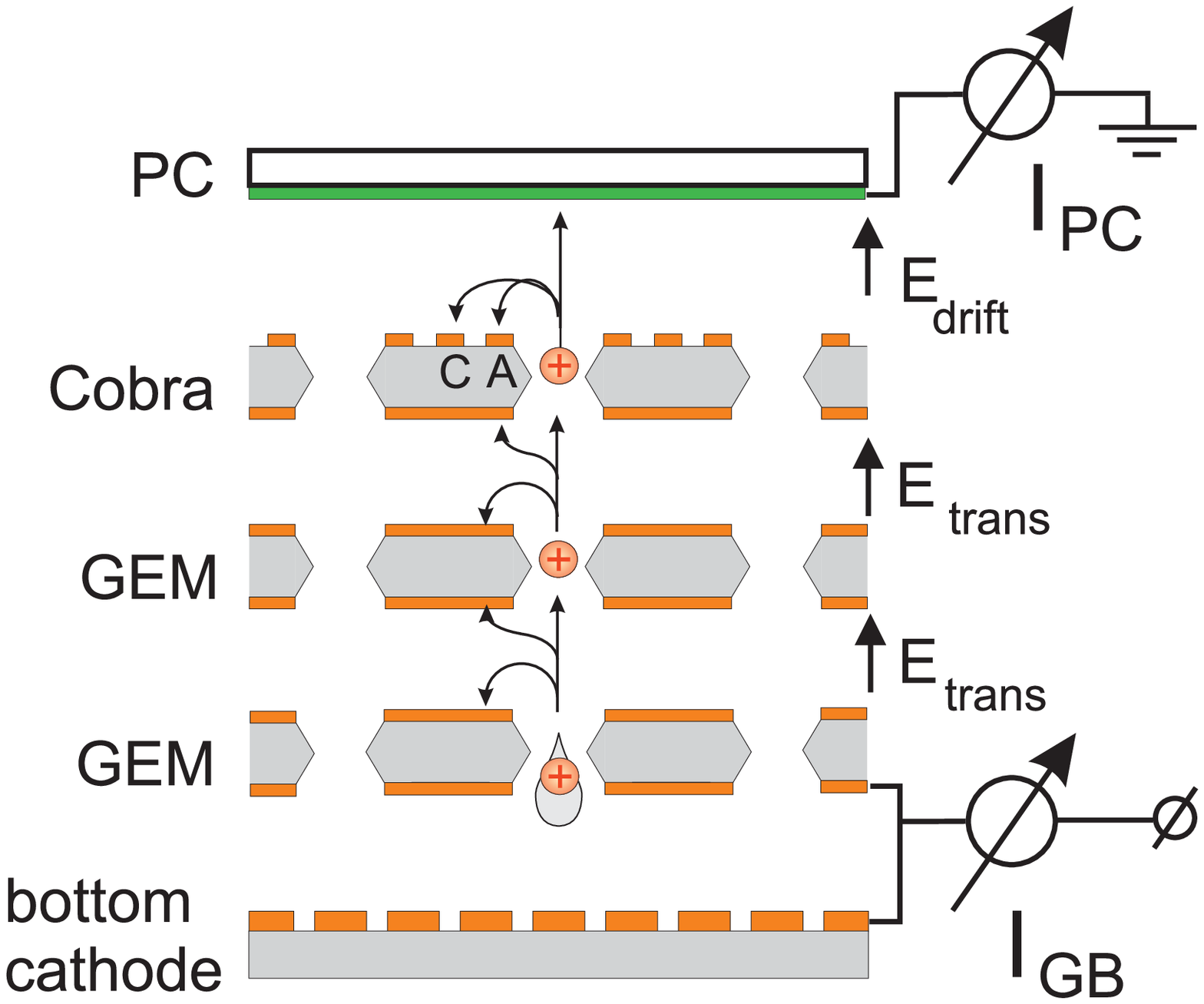}
}\hspace{0cm} %
\subfigure [] [] 
{
    \label{fig:4:b}
    \includegraphics[width=7cm]{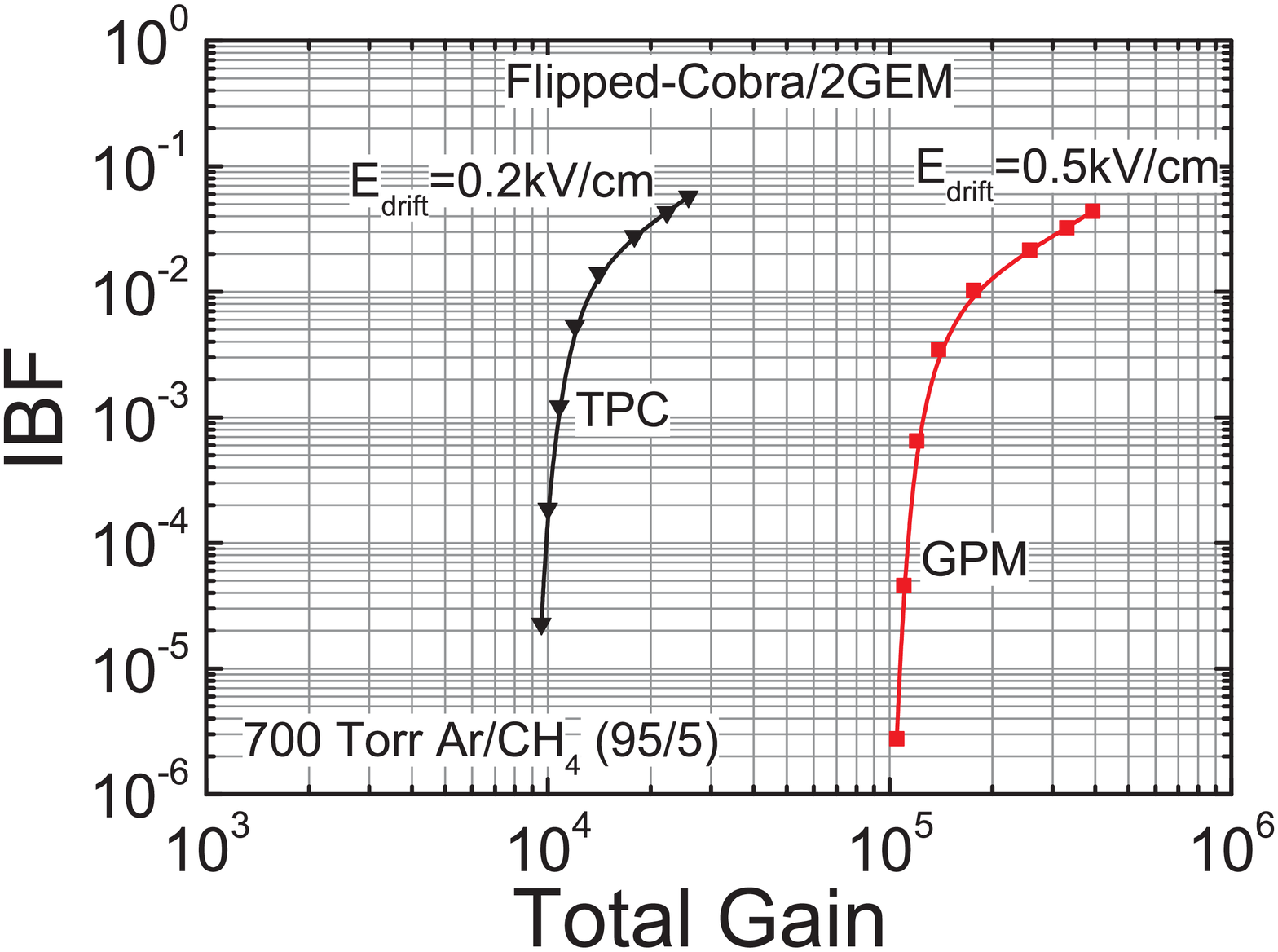}
}%
\caption {(a) Scheme of cascaded Cobra/2GEM GPM with a
semi-transparent photocathode; possible avalanche ions paths are
also shown. (b) The IBF as a function of the total gain of the
Cobra/2GEM cascaded detector for drift fields of 0.2 kV/cm (TPC
conditions, triangles) and 0.5 kV/cm (GPM conditions, squares).}
\label{fig:4}
\end{center}
\end{figure}

Following the success of the above study, and with our understanding
of the operation mechanism of the MHSP-like electrodes
\cite{lyashenko:07}, a new patterned micro-hole electrode named
Cobra (\ref{fig:3}) was developed with a geometry that is expected
to improve the ion divergence away from the holes. It has thin anode
electrodes surrounding the holes and creating strong electric field
inside the holes (required for charge amplification); the more
negatively biased cathode electrodes cover a large fraction of the
area for better ion-collection as compared to the F-R-MHSP
\cite{lyashenko:07}. The concept of the Cobra electrode has been
recently investigated. It was found that when introduced as a first
element (with the patterned area pointing towards the photocathode),
preceding two GEMs in the cascade (\ref{fig:4:a}), it drastically
improved the ion trapping capability. The IBF as a function of
voltage between electrodes on the top surface of Cobra is shown in
\ref{fig:4:b}. In GPM conditions, with a drift field of 0.5kV/cm, we
measured IBF values of $3\cdot10^{-6}$ which is 10,000 times lower
than that of cascaded triple GEMs. In TPC conditions with a drift
field of 0.2kV/cm the same detector configuration provided IBF
values as low as $2.7\cdot10^{-5}$. These are the lowest IBF values
ever reached in gaseous detectors. However, while the F-R-MHSP
yielded full photoelectron collection efficiency into the holes of
the first cascade element, the Cobra, in its present geometry, had a
limited electron collection efficiency of about 20\%. This can and
should be improved by optimizing the geometrical parameters.

\section{DC operation of a visible-sensitive GPM with micro-hole
multiplier cascades}

\begin{figure} [h]
  \begin{center}
  \makeatletter
    \renewcommand{\p@figure}{Fig.\space}
  \makeatother
    \epsfig{file=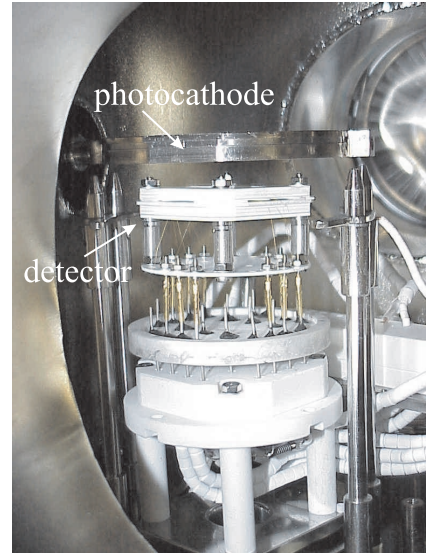, width=5.5cm}
    \caption{Photograph of the F-R-MHSP/GEM/MHSP detector and the photocathode, mounted in the vacuum chamber.}
    \label{fig:5}
  \end{center}
\end{figure}

The operation of a visible-sensitive GPM in DC mode was investigated
with a F-R-MHSP/GEM/MHSP cascaded multiplier, schematically shown in
\ref{fig:2:a}. A photograph of the experimental detector is shown in
\ref{fig:5}. All the multiplier electrodes were mounted between
ceramic spacers within a UHV vessel \cite{moermann:thes}. The
photocathode was prepared and characterized in an adjacent vessel of
the dedicated UHV system, and then transported with an UHV
manipulator and placed in a stainless-steel holder above the
detector. Details can be found in \cite{balcerzyk:03} and in
\cite{moermann:thes}. The K-Cs-Sb PCs had typical quantum efficiency
(QE) of 30\% measured in vacuum at 375 nm.

The details of the IBF measurements and results for this multiplier
configuration were reported in \cite{lyashenko:07} both in
conditions for TPC and for GPM operation (\ref{fig:2:b}). Conditions
for full efficiency of electron collection from the PC were
confirmed and applied in all measurements

\begin{figure} [h]
  \begin{center}
  \makeatletter
    \renewcommand{\p@figure}{Fig.\space}
  \makeatother
    \epsfig{file=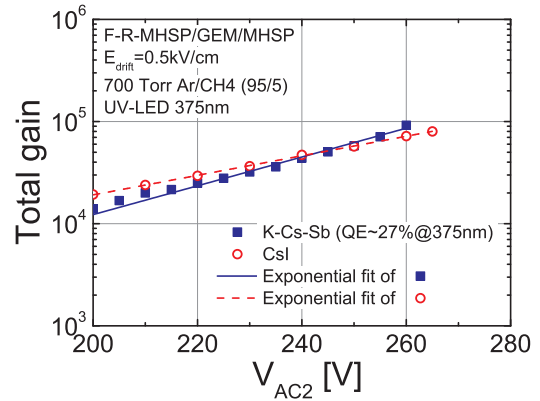, width=7cm}
    \caption{Gain-voltage characteristics of the detector shown in
\protect\ref{fig:2:a} with a K-Cs-Sb (squares) and CsI (circles)
photocathodes. The data was fitted with exponential functions; no
divergence from exponential was observed. 700 Torr Ar/CH$_{4}$
(95/5); E$_{drift}$=0.5 kV/cm. QE refers to vacuum.}
    \label{fig:6}
  \end{center}
\end{figure}

In \ref{fig:6} we present gain-voltage characteristics for the
cascaded GPM of \ref{fig:2:a} with a K-Cs-Sb PC and with a CsI PC,
for comparison. The measurements were carried out in Ar/CH$_{4}$
(95/5) at 700 Torr. The present semitransparent K-Cs-Sb PC had a QE
of $\sim$27\% measured in vacuum at 375nm. Its QE value in the gas,
with a drift-field of 0.5 kV/cm, was estimated to be 16\%
\cite{moermann:thes}. The solid and dashed curves in \ref{fig:6}
represent exponential fits to the data points measured with K-Cs-Sb
and CsI PCs, correspondingly. In both cases the GPM could reach a
gain of 10$^{5}$ with no divergence from an exponential gain-voltage
characteristic, indicating upon full suppression of ion feed-back
effects.

\begin{figure} [h]
  \begin{center}
  \makeatletter
    \renewcommand{\p@figure}{Fig.\space}
  \makeatother
    \epsfig{file=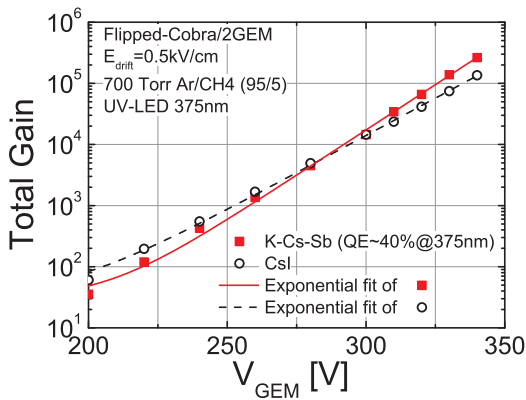, width=7cm}
    \caption{Gain-voltage characteristics of the Cobra/2GEM cascaded GPM
of \protect\ref{fig:4:a}, with K-Cs-Sb (squares) and CsI (circles)
photocathodes. The data were fitted with exponential functions. 700
Torr Ar/CH$_{4}$ (95/5); E$_{drift}$=0.5 kV/cm. QE refers to
vacuum.}
    \label{fig:7}
  \end{center}
\end{figure}

The visible-sensitive GPM with a K-Cs-Sb PC coupled to the Cobra
multiplier followed by two GEMs (\ref{fig:4:a}) was investigated in
DC operation mode; the gain-voltage plots are shown in \ref{fig:7},
in comparison with a CsI PC. The measurements were carried out in
Ar/CH$_{4}$ (95/5) at 700 Torr. The semitransparent K-Cs-Sb PC had
$\sim$40\% QE measured in vacuum at 375nm. The exponential fits to
the data points measured with K-Cs-Sb and CsI PC are represented by
solid and dashed curves, correspondingly. There were no feed-back
effects as can be seen from the exponential shape of the
gain-voltage curve.

\section{Conclusions}

Ion feedback in cascaded micro-hole gaseous detectors was studied
with a variety of cascade elements and PCs, in conditions of TPC and
of GPM operation. The effective ion feedback probability
$\gamma_{+}^{eff}$ was measured in Ar/CH$_{4}$ (95/5) at 700 Torr
and found to be $3\cdot10^{-2}$ for Na-K-Sb, K-Cs-Sb and Cs-Sb
photocathodes. Based on these measurements, the ion backflow
fraction (IBF) required for stable DC operation of cascaded
visible-sensitive gaseous photomultipliers (GPM) was estimated to be
$3.3\cdot10^{-4}$.

Systematic ion blocking investigations with various patterned
micro-hole cascaded multipliers yielded the required IBF values, at
gain of $10^{5}$. The best results were recorded in a cascaded
multiplier of a flipped reversed-bias micro-hole and strip plate
flowed by a GEM and by a micro-hole and strip plate
(F-R-MHSP/GEM/MHSP). This configuration yielded 100 fold lower IBF
value than that measured in cascaded GEMs. This permits reaching
stable operation conditions both for TPCs and for visible-sensitive
GPMs operating in DC mode.

Even lower IBF values, of $3\cdot10^{-6}$ at a gain of $10^{5}$ and
drift field 0.5kV/cm, was recorded in a cascade comprising a novel
"Cobra" micro-hole patterned multiplier, followed by two GEMs. This
IBF value is 10,000 times lower than that measured in cascaded GEMs.
However, the electron collection efficiency of the present "Cobra"
multiplier was only 20\%, which requires further optimization of its
geometry.

A Visible-sensitive GPM with a F-R-MHSP/GEM/MHSP cascaded multiplier
and a K-Cs-Sb photocathode, yielded, for the first time, stable
operation at gains of $10^{5}$ in DC mode with full photoelectron
collection efficiency and without any noticeable feedback effects.
This is a breakthrough in the field of gaseous photomultipliers

\section*{Acknowledgments}
This work is partly supported by the Israel Science Foundation,
grant No 402/05, by the MINERVA Foundation and by Project
POCTI/FP/63962/2005 through FEDER and FCT (Lisbon). A. Breskin is
the W.P. Reuther Professor of Research in The Peaceful Use of Atomic
Energy.

\bibliographystyle{elsart-num}
\bibliography{publications}

\end{document}